\documentclass{article}
\pdfoutput=1

\usepackage{arxiv}

\usepackage[utf8]{inputenc}
\usepackage[T1]{fontenc}
\usepackage{hyperref}
\usepackage{url}
\usepackage{booktabs}
\usepackage{amsfonts}
\usepackage{nicefrac}
\usepackage{microtype}
\usepackage{graphicx}
\usepackage{doi}
\usepackage[margin=0.5cm]{subcaption}
\usepackage{float}

\RequirePackage[labelfont={bf},%
                labelsep=period,%
                justification=raggedright]{caption}

\begin{document}
\title{A Methodology for a Scalable, Collaborative, and  Resource-Efficient Platform to Facilitate Healthcare AI Research}

\author{ \href{https://orcid.org/0000-0002-2251-6740}{\includegraphics[scale=0.06]{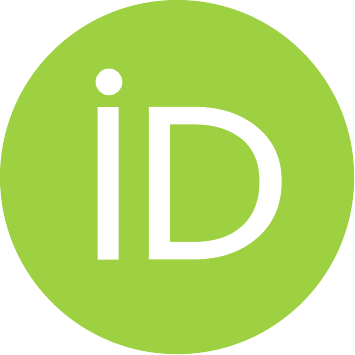}\hspace{1mm}Raphael Y.~Cohen}\\
	Department of Anesthesiology, \\
	Perioperative, and Pain Medicine\\
	Brigham and Women's Hospital\\
	Boston, MA 02115 \\
	\texttt{rcohen15@bwh.harvard.edu} \\
	\And
	\href{https://orcid.org/0000-0001-7367-3394}{\includegraphics[scale=0.06]{orcid.pdf}\hspace{1mm}Vesela P. ~Kovacheva} \\
	Department of Anesthesiology, \\
	Perioperative, and Pain Medicine\\
	Brigham and Women's Hospital\\
	Harvard Medical School\\
	Boston, MA 02115 \\
	\texttt{vkovacheva@bwh.harvard.edu} \\
}

\renewcommand{\shorttitle}{A Methodology for a Scalable, Collaborative, and  Resource-Efficient Platform to Facilitate Healthcare AI Research}

\hypersetup{
pdftitle={A Methodology for a Scalable, Collaborative, and  Resource-Efficient Platform to Facilitate Healthcare AI Research},
pdfsubject={eess.IV, cs.AI, cs.DC},
pdfauthor={Raphael Y.~Cohen, Vesela P.~Kovacheva},
pdfkeywords={Electronic Health Records, Artificial Intelligence, Distributed Systems},
}

\maketitle

\begin{abstract}
Healthcare AI holds the potential to increase patient safety, augment efficiency and improve patient outcomes, yet research is often limited by data access, cohort curation, and tooling for analysis. Collection and translation of electronic health record data, live data, and real-time high resolution device data can be challenging and time-consuming. The development of real-world AI tools requires overcoming challenges in data acquisition, scarce hospital resources and high needs for data governance. These bottlenecks may result in resource-heavy needs and long delays in research and development of AI systems. We present a system and methodology to accelerate data acquisition, dataset development and analysis, and AI model development. We created an interactive platform that relies on a scalable microservice backend. This system can ingest 15,000 patient records per hour, where each record represents thousands of multimodal measurements, text notes, and high resolution data. Collectively, these records can approach a terabyte of data. The system can further perform cohort generation and preliminary dataset analysis in 2-5 minutes. As a result, multiple users can collaborate simultaneously to iterate on datasets and models in real time. We anticipate that this approach will drive real-world AI model development, and, in the long run, meaningfully improve healthcare delivery.
\end{abstract}

\section{Introduction}

Recent advances in artificial intelligence (AI) in healthcare hold the potential to increase patient safety, augment efficiency and improve patient outcomes. In clinical care, AI technologies can aid physicians in diagnosis and treatment selection, risk prediction and stratification, and improving patient and clinician efficiency \cite{He2019ThePracticalImplementation}.  There has been a vast expansion of available AI technologies in the past decade, creating considerable interest in healthcare data science. Yet, most sophisticated AI models exist only in high-profile publications, and only a few are implemented in clinical practice \cite{Panch2019, Chen2019HowToDevelopML}.

The barriers to translating data science research into patient care are inadequate data quality, scarce resources, and high patient confidentiality needs. With the Health Information Technology for Economic and Clinical Health Act of 2009, many institutions have transitioned to electronic medical records that provide a rich medical data source. While initially developed for administrative purposes, most electronic health record (EHR) systems store patient data in heterogeneous formats, sometimes combined with legacy systems. In addition to the structured data for medications, laboratory data, and imaging, there are large amounts of unstructured data like physician notes, discharge summaries, and reports. The EHR data has a significant degree of missingness, misclassification, and errors \cite{2020ReviewofChallenges}. Furthermore, high-dimensional data like telemetry, EEG, genetic, and continuous physiological data provide valuable clinical insights but require large storage and processing capacity. Harmonizing these diverse data sources can be a very time-consuming and resource-heavy process. In contrast to the large technology companies, most healthcare entities lack interoperability of their data sources, high compute resources, and large data science teams. Moreover, healthcare is a highly regulated industry, and there are complex requirements for patient safety, confidentiality, and ethics. These characteristics of the healthcare landscape present significant challenges for the rapid and efficient development and deployment of AI technologies.

\begin{figure}[ht]
\centering
\includegraphics[width=0.9\linewidth]{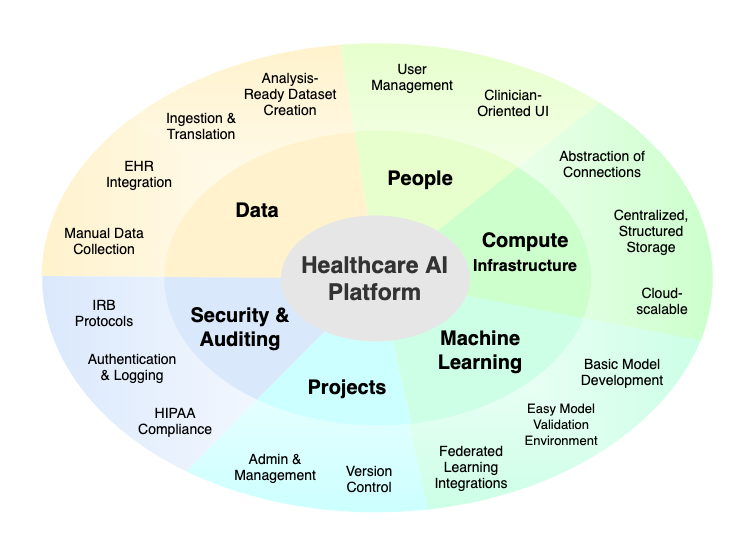}
\caption{Broad overview of functions and capabilities enabled by our healthcare AI platform.}
\label{fig:merlin_functions}
\end{figure}

We aimed to create a system that can automate clinical data acquisition and processing efficiently in machine-learning-ready formats, enabling near-real-time generation of datasets for subsequent analysis and development of AI models. At the same time, scalability to overcome significant data needs and generalizability to handle heterogeneous projects are requirements for this system. In addition, we followed the strict data governance and patient privacy requirements in our institution. Achieving these objectives yields a low-friction platform that focuses on iterative collaborative processes, making the AI workflow more productive and efficient. The functionality of the platform we created is shown in Figure \ref{fig:merlin_functions}. We demonstrate that attention to early challenges with data translation can offer substantive benefits at later stages of dataset and model creation. This platform will significantly reduce time to model completion and thus create opportunities for faster AI integration in clinical care to meaningfully impact patient outcomes.

\section{Related Work}
Multiple systems exist that can automate data collection and warehousing; recent works demonstrate progress in data translation and the AI model lifecycle.

Large research patient data registries can store and provide research data access efficiently. Those entities can store diverse data from cohorts of hundreds of thousands of patients that originate from EHRs, genomics, surveys, and personal devices; for example, i2b2 \cite{Murphy2010ServingTheEnterprise}, UKBiobank \cite{Sudlow2015UKBiobank}, and AllofUS \cite{Denny2019AllOfUS}. As a result, large teams with substantial informatics expertise are needed to maintain and provide services to the research community. Unparalleled opportunities to explore vast amounts of data have created new research endeavors. However, these large repositories were not developed primarily for AI research and modeling, and further processing of data is often required. Research teams are expected to be equipped with substantial technical expertise to clean the data and develop specific use-case models. In addition, these systems are not nimble, and once a dataset is created, adding more variables may necessitate submitting a new request, developing new data processing pipelines, and re-running all downstream processing and modeling components.

Multiple computer systems exist to automate the data collection process \cite{2015SecondaryUseOfElectronic, 2018ScalableAndAccurate, 2019UnravelBigData, Arora2019Isthmus, 2020HealtheDataLab, 2021CombiningStructured}. Extract, Transform, Load (ETL) pipelines using clinician-informed rules \cite{2013LeveragingElectronicHealthcare, 2016HybridSolutionForExtracting, Khurshid2021CohortDesign} or natural language processing (NLP)-aided systems \cite{2018ArchitectureAndImplementation, 2020AugmentedCuration} have demonstrated success in curating data warehouses for targeted use cases. Due to differences in nomenclature and notation used to document hospital data over many years and changes in clinical practice, these solutions may fail to translate to other institutions and use cases \cite{2019Heterogeneity}.

In addition, updating or modifying those pipelines may require significant expertise and resources. Some of these challenges in data harmonization can be overcome using atomic data attribute representations in a data warehouse, in which the data are stored in its most granular form \cite{Visweswaran2021AnAtomicApproach}. This approach allowed resource- and time-efficient data management.

A recent platform for medical imaging AI utilized a microservice architecture with an emphasis on early and accurate translation of data to machine amenable formats to lower the barriers to entry for AI development \cite{Tomosuite}. This platform reduced data acquisition, cohort development, and model training time by enabling continuous development of both data cohorts and models.

Instead of limited use-case platforms or model-dependent ETL processes, we sought to create a framework agnostic to the data sources, which can be implemented to serve the needs of a wide variety of projects.

\section{Bottlenecks in the Biomedical Data Analysis Workflow}
Solving challenges that occur late in the workflow by addressing them in earlier stages enables the platform's scale and efficiency. This allows for simple algorithms to accomplish formerly complex needs within the workflow. The stated goals for the platform will be distilled into three measurable objectives as pertaining to biomedical data analysis; then the principles for handling both data and algorithms that govern the platform  design will be presented; and finally we will show how we applied those principles to implement the platform and answer our established objectives. 

\begin{figure}[ht]
\centering
\includegraphics[width=\linewidth]{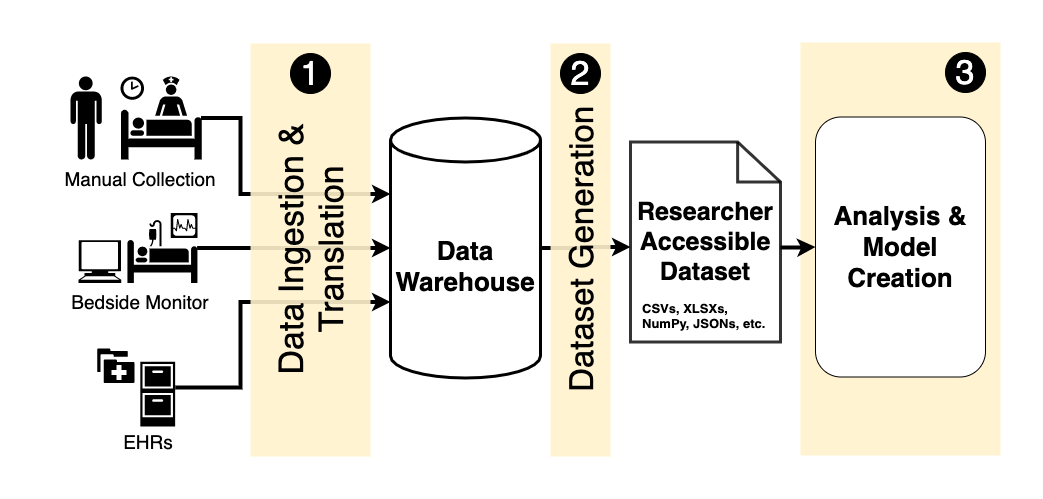}
\caption{A general high-level workflow through biomedical data analysis. Areas most benefiting from optimization are highlighted as 1. The data ingestion and translation process, 2. The dataset generation process, which is often iterative as phenotypes and variable are selected and refined, and 3. The analysis process itself.}
\label{fig:general_method}
\end{figure}

Biomedical data analysis generally follows the framework shown in Figure \ref{fig:general_method}. Data is acquired and ingested, stored, and then used for analysis.

The challenges associated with healthcare data have been extensively discussed \cite{2020ReviewofChallenges} and include missing data, redundant data, lack of structure and standardization. For example, "preterm", "Pre-Term", and "premature" all refer to the same concept, preterm delivery, despite appearing differently within the EHR. Likewise, the process of ingesting data can be time consuming and resource intensive when new, often unstructured, data becomes available for use on new projects. Poor data standardization from source data can propagate into the data warehouse and beyond into subsequent datasets. As a result, such patient cohorts would need to be verified due to the risk of variable representations within derived values. Furthermore, downstream modeling thus becomes more laborious due to those data irregularities. To reduce many of these bottlenecks, our platform focuses on addressing later challenges by minimizing their impact during earlier stages of the workflow.

Figure \ref{fig:general_method} shows the three highlighted areas of interest where this propagation can be minimized. Our platform addresses these in three objectives, which will be used as touchstones in analyzing and evaluating the platform:

\begin{itemize}
    \item Objective 1: Acquire, ingest and standardize data automatically into a machine-analysis-ready form.
    \item Objective 2: Generate data cohorts efficiently and comprehensively. This can enable researchers to work dynamically with diverse collaboration. 
    \item Objective 3: Generate standardized datasets in a format for common analysis packages, such as NumPy, CSVs, Excel, etc. This dataset creation process should occur in near-real-time.

\end{itemize}

Additional considerations include the amount of data, which may be in the hundreds of thousands of patients, the security and regulatory needs for the platform's execution, and the management of people and projects within the system so that the created tools can serve as an interactive workbench for researchers.

\section{Principles for Data Handling and Algorithm Design}

Principles that govern the development of the platform are divided into principles for the handling and storage of data, and principles for the development and organization of algorithms.

\subsection{Principles for Data Storage}

\begin{itemize}
    \item \textbf{Atomicity/Indivisibility of Data Attributes}: Once data are translated, it must be stored as data attributes that cannot be broken down into meaningful sub-components. For example, "Test Result": "Positive covid19 antibody test" can be broken down into sub-components, whereas each item in the collection of "Test Type": "covid19", "Test Subtype": "antibody", "Test Result": "positive" cannot be further broken down into meaningful sub-components.
    \item \textbf{Consistency}: The same value should have the same meaning across the entire dataset, and multiple values cannot be used to represent the same concept. For example, "Positive", "Pos", "True", and "Yes" are all results for a covid19 test that indicate a positive result. Instead of this variety in presentation, the system instead stores the boolean value True to indicate the test was positive, False if it was negative, and Null if it was inconclusive.
    \item \textbf{Translatability}: Data that are not stored as numeric values must have a complete mapping or methodology provided to convert these data to a numeric format. Data that are stored numerically may need a mapping to a human-readable value. For example, "low", "medium", "high" might map to 0, 1, 2.
    \item \textbf{Traceability}: Data must be tracked through its entire lifecycle within the system. Data that has been translated must maintain a reference to its source data. Data that exists in a version-controlled dataset must maintain a reference to the translated data from which it was drawn. All models must maintain a reference to the dataset it was developed with.
\end{itemize}

\subsection{Principles for Modular Algorithm Development}

Every component of the platform architecture is a containerized module that performs simple and small-scale individual tasks without cross-module dependencies. This reduces bottlenecks and enables parallelization that can take advantage of cloud-computing resources and distributed workloads. At the core of the system are common shared resources:

\begin{itemize}
    \item \textbf{PostgreSQL Database}: Storage of most data in a relational database, including raw EHR, translated data, manual data collection, projects, management records.
    \item \textbf{Simple Storage Service (S3)}: High-resolution data, large text files, version-controlled datasets, and models are stored in an S3 resource. This resource provides file-like storage and is common in services like Amazon Web Service (AWS).
    \item \textbf{Messaging Queue Service (MQ)}: 
    A service which allows worker modules to post and read jobs to and from queues like a messaging board. When modules complete their tasks, they inform the message queue resource the job is complete, and they receive the next job to work on. Many modules, especially copies of the same module, may all operate on the same queue at once to provide large-scalability of processing capacity.
    \item \textbf{Networking Services}: Some modules serve user-facing applications and need to communicate with other networked devices, so they interact with the stack of network services including NGINX to connect to their endpoints.
\end{itemize}

\begin{figure}[ht]
\centering
\includegraphics[width=0.9\linewidth]{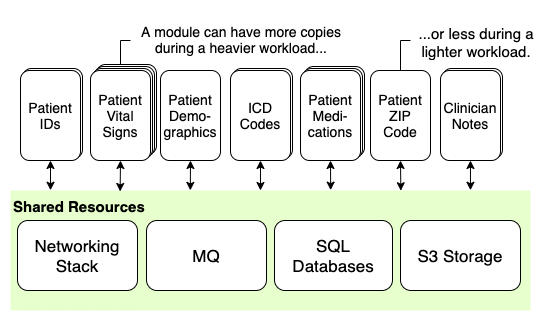}
\caption{Worker modules communicate only with core shared and stateful resources. Worker modules are stateless, and the number of copies of each type can be scaled up and down on-the-fly to match the real-time load on the platform. Note that the modules shown are not exhaustive of all modules in the platform.}
\label{fig:merlin_parsing}
\end{figure}

On top of this layer are hundreds of worker modules, which are stateless services that can only communicate with the MQ resource and the shared storage resources, but not with each other. This is shown in Figure \ref{fig:merlin_parsing}. These modules run in virtualized Docker containers, and can run on distributed compute systems so long as they can communicate with the core resources. Functions that a user of the system may view as a single task, such as pulling data for specific patients from the EHR, creating datasets, or running basic statistics are not handled by a single module, but rather by a collection of many modules, often in parallel, before the results are returned to the user. Each module performs small, simple parts of a workflow that in aggregate performs larger tasks. For example, one module's function may be to analyze a request to pull data from the EHR, and split it into appropriately sized requests for another module that actually pulls the data from the EHR. Another module might upload the results from the EHR into a staging database, before additional specialized modules process that data. This design enables an ensemble behavior that allows an automated system controller such as Kubernetes to dynamically scale copies of modules with the immediate computational needs within the platform, swarming large jobs with temporary resources at appropriate scales.

The centralization of the network stack allows the system to maintain one path in and one path out, thereby simplifying efforts needed to maintain security and logging, as well as to meet regulatory needs.

\section{Applying Principles to Create an Interactive Workbench Framework}
\label{methods}
The implementation of these principles is reflected throughout the development of this platform and in the various workflow components. For each of the following workflow sections, modules are developed that allow developers to "plug-and-play" with specific needs for new sources; module templates serve as a framework that allows this platform to be readily adapted to new data, sources, and needs.

For example, in the data ETL modules, there is a core processing component unique to each source (such as an EHR, legacy database, or live data collection) which is contained in a common shared framework. Including a new data source is therefore as simple as using this framework to duplicate a template, and add the custom code needed to interpret the new source. 

The core layer of the platform is the data collection, translation, and analysis pipeline. On top of this core layer are layers for management and organization, security, and applications to interact with this environment. Because this platform focuses on addressing early needs in the pipeline, the methods in the core layer will first be discussed in the order that data flows through it, followed by the methods included in the layers on top of it.

\subsection{Data Collection Processes}

\begin{figure}[ht]
\centering
\includegraphics[width=0.8\linewidth]{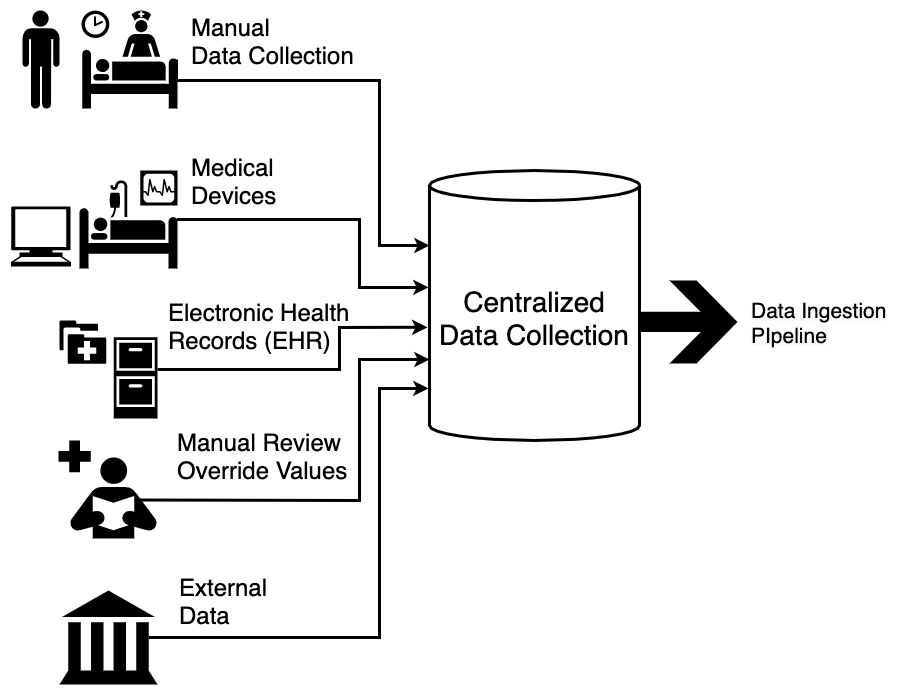}
\caption{Multimodal data sources feeding into a centralized data collection process in the platform.}
\label{fig:merlin_sources}
\end{figure}

Data can be collected from multimodal sources. The platform includes pathways for real-time and retrospective data collection. Tools were created to enable bedside structured data entry, which along with a timestamp, are continually streamed back to the server application. Shown in Figure \ref{fig:merlin_sources} are also software wrappers that encapsulate medical device recording software and stream continuous waveform data, such as 300Hz ECG data, back to the server.

The platform also supports querying and pulling data from the hospital EHR infrastructure. Given the depth of the EHR data, these sources also represent a large trove of longitudinal data on vast retrospective patient cohorts. There are sometimes errors with data stored in the EHR, and the platform includes a QA/QC pipeline to detect and mitigate many of these errors. Any manual corrections obtained during the QA/QC process are recorded and treated as an additional data source.

Additionally, for collaborative efforts, external data may be uploaded and stored. These data will be processed and transformed into the same standardized formats as internal data.

Every data resource has its own data source controller, which can initiate a query (for resources such as the EHR) and receive data, including streaming capability in the case of medical device recording.

These data are recorded and stored in S3. To avoid bottlenecks in case of large tasks, the  data are first stored before being processed to separate out the operation to receive data from its interpretation. When the data is ready to be processed, these controller modules publish a message to a messaging queue to inform the collection of data upload controllers to ingest the files into the proper locations for the subsequent data ETL modules. This upload controller receives the instructions from the message queue informing it what type of data was received, and where its location for processing is, such as in the source database for most data, or reformatted data in S3 for waveform analysis. Once the data is present, it creates a series of messages in the appropriate message queues pointing to small blocks of unprocessed data.

Once data has passed through this stage, it is ready to be translated and interpreted.

\subsection{Data ETL Processes}
\label{etlmodules}

To facilitate the data translation process, a python library was developed. This library standardizes the creation of stateless Docker modules which handle all communication with the message queue, acquisition of source data, and the insertion and conflict resolution for ingested data. For each new data item, a developer need only define the internal logic in a single function which translates a single row of source data into a list of output data elements.

ETL modules adhere to the data principles established above. Output data is consistent in representation (numeric or decimal is preferred, short text if required), in unit (standardized to the metric system), and if text data, in choices. Concepts that are typically stored as long strings in the source data, such as the formulation of drugs in the EHR, are broken down into their components. For example, the value "BUTALBITAL-ASPIRIN-CAFFEINE 50 MG-325 MG-40 MG CAPSULE" is broken down into three linked entries (medication class, dose, units): ('aspirin', 325, 'mg'), ('butalbital', 50, 'mg'), and ('caffeine', 40, 'mg'). These three entries are trivially reassembled to indicate that the patient received these three drugs at once.

All modules are designed with medically-relevant knowledge stored in JSON configuration files. This decreases required maintenance efforts for developers and allows for the creation of interfaces that let clinicians update modules without the need for programming experience. If data is found that can not be translated with the framework in the configuration file, it is flagged for QA/QC review. Furthermore, if data is translated but produces unexpected results, it is automatically sent to the QA/QC pipeline, and may be withheld based on the severity of the deviation.

Modules are specialized to small and specific ETL tasks. For example,  one module handles ICD code extraction from EHR data, another handles vital signs from the EHR flowsheet data, and another processes waveform data from live data collection. This is done so that Kubernetes can dynamically scale the number of each of these modules with the amount and kind of data waiting to be translated. This is also shown in Figure \ref{fig:merlin_parsing}. The types of data ETL modules presented are merely exemplary in the figure, and the entire system contains many dozens of ETL modules based on the data form and type.

With the exception of data passed to the QA/QC pipeline, once data have moved through this phase, they are stored in the data warehouse in adherence with the outlined data principles in order to address Objective 1.

\subsection{Quality Assurance and Quality Control Processes}

The QA/QC pipeline begins with one of the following pathways.

\begin{itemize}
    \item \textbf{Errors in the ETL Pipeline}: Triggered when data cannot be parsed in the ETL processes. This includes the case when a processed value lies outside of an expected range.
    \item \textbf{Data Warehouse Validation}: Data is selected at random to ensure the ongoing quality of the data warehouse.
    \item \textbf{Dataset Validation}: A user specifically requests validation for a generated dataset.
\end{itemize}

\begin{figure}[ht]
\centering
\includegraphics[width=0.8\linewidth]{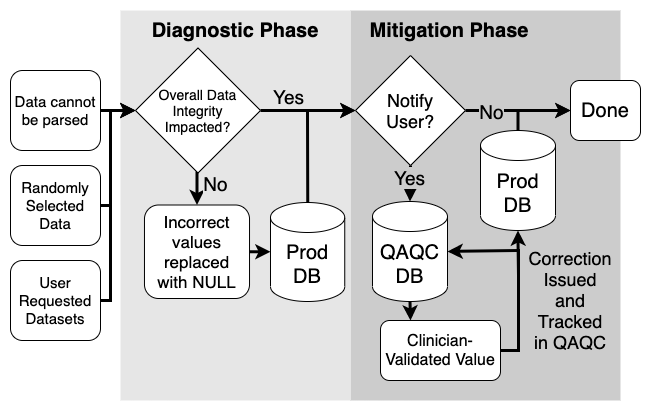}
\caption{The quality assurance process which consists of a diagnostic phase and a mitigation phase. Data is quarantined until an expert is able to provide mitigation.}
\label{fig:merlin_qaqc}
\end{figure}

The majority of the QA/QC processes pertains to the first pathway. If the integrity and accuracy of a data point is in question, the ETL modules carry out a diagnostic process on the data and stage it for manual mitigation, shown in Figure \ref{fig:merlin_qaqc}.

Once data is moved to the QA/QC pipeline, the mitigation phase begins. Data is inserted into a special QA/QC database, in different tables based on its source. Data coming from the ETL process maintains links to the source data in question along with a status message from the ETL module describing the source of problem.

The clinician uses available tools, such as their access to the EHR along with their medical background to decide what the correct input or parsed value should have been. They can submit a correction to the source data, which is used as a secondary source of input that overrides the source data. In addition to the correction, the physician's sign off and timestamp are recorded by the system.

If a QA'ing physician believes that an ETL module is producing a systematic error, they can issue a HALT command to terminate those modules. Once developers have upgraded the module, all data processed by that module will be re-processed.

In the case of a QA/QC process on a dataset, the results of the operation are also stored with the dataset.

\subsection{Dataset Creation Processes}

The workflow components above address Objective 1. The benefits of early attention to data quality and translation can now be realized as Objective 2 is addressed.

Dataset creation is designed to be an interactive process for researchers. Researchers are able to define complex variables and custom phenotypes to create datasets from the atomically stored data elements in the data warehouse. Variables are defined as a

\begin{itemize}
    \item Data Source: Either a table and field, or a set of custom variables
    \item Operation: A function to apply to the data. It may be as simple as "Identity" to return the existing value, more complex like "Like", which is similar to the SQL Like operator, or even "Time Series" to aggregate entire series with timestamps, values, and dates. Numerous operations are defined within the system, and developers can contribute to a library available for all users.
    \item Value: An externally provided value or list of values to be used with the Operation. For example, 'O10.' might be provided to the "Right Like" operator to yield all ICD codes that begin with 'O10.'.
    \item Timeframe: A limit on when the operation can be applied. These can be patient dependent, such as "During Pregnancy" or "Before Visit", or during specified time frames, such as after January 1, 2015.
    \item Constraints: A list of other variables to use as constraints on the current operation.
    \item Mapping Scheme: An optional choice that lets users define how data can map to numeric values. The default is stored mappings that are created at the same time as translation modules are developed, but users may choose from common orderings, alphabetical order, or create their own.
\end{itemize}

When the user launches the dataset generation process, a series of modules extract the relevant information from the data warehouse with a series of parallel SQL commands and data stores in S3. The commands are optimized to limit both the complexity of the queries on the database, and the size of the data being extracted from it. It then compiles the dataset into a single CSV or NPY file available for further use in the application or user download. Two copies of the dataset are stored: One is a human-readable document with standardized strings and text, and the other is machine-analysis ready, which uses the mapping scheme for each variable to create a numeric dataset. Both datasets are version-controlled with additional references and logging stored in a database table. 

Following the creation of the dataset, the next phase, dataset analysis, is automatically triggered.

\subsection{Dataset Analysis Processes}

The presence of machine-readable datasets simplifies the automation of basic statistics, with the target of satisfying Objective 3. A series of modules work in concert to create dataset summary and basic statistical analysis upon dataset creation. Users of the system can define additional statistics, or discrete or continuous groups within the data for more advanced summary and statistics usage.

The dataset summary includes mean, median, standard deviation, extreme values, and percentile distributes or each variable and group. These data can be used to create graphs within software applications. The statistics module creates correlation maps in addition to user-specified statistical tests. These include t-tests, Pearson and Spearman correlation, and ANOVA tests. All results are user exportable in CSV files.

When researchers are satisfied with a dataset, they can use modules within the platform to train many deep learning models in parallel. The built-in model zoo includes FCDNNs, CNNs, LSTMs with Attention Mechanism, Transformers, and other common models. Model checkpoints are saved and researchers can track training progress. Users may also download generated Python code and model weights to further adapt the trained models. Models may also be exported as Docker images for use in applications.

This completes the data collection, translation, and analysis core layer. In order to operate within a HIPAA-compliant environment, and to facilitate interaction with both data and the workbench environment, additional layers are needed.

\subsection{Security and Integrity Layer}

All actions within the platform are timestamped and logged. All data is stored with a linking identifier back to its source as well as the version of the module used to translate it. All datasets are version controlled and maintain links to the versions of the modules used to translate the data from which it was derived. Collectively, this allows data at any stage in the pipeline to be traced to the source data it originated from and what code processed it along the way. Combined with the logs of user activity and limits and permissions for users to perform necessary functions, this contributes to the system's design for accountability and integrity.

All activity in and out of the platform is further logged by the system. In order to provide IRB and HIPAA compliant operation, the system enforces that all data is associated with IRB protocols, and that projects, researchers, and data (or datasets) all must belong to the same IRB protocol in order for them to be accessed. User access can be controlled by individuals with administrative privileges.

Combined with logs, this can facilitate auditing and regulatory needs for the platform.

\subsection{Application Layer}

Numerous applications make up the platforms' ecosystem. These range from data-collection oriented tools, research tools, and even management applications.

Data collection applications operate locally on computers for bedside use. Following authentication, users select an appropriate project, and create and update entries for patients. The data generated by the tool are structured and standardized. Structured data entry types can be created by authorized users on a centralized web client application.

Traditionally, in live recording sessions, users may not be able to record all aspects of an event in the exact moment it occurs. Errors in event timestamps may arise from recording delays. To minimize this error, in the platform's data collection applications, users press a button which immediately records the timestamp of the event. They then are able to fill in any supplemental numerical values at the first available moment that they can safely do so.

These applications store temporary anonymized backups of data collection locally in the event of a network outage, and opportunistically stream data to the server when network connectivity returns. The interface of this application was designed to maximize usability and efficiency for bedside use.

The data collection applications also integrate with other solutions that record live data from medical devices and are able to initiate and control streams back to the platform. Other desktop solutions designed for data collection can be integrated as well, and some may get software wrappers or daemon applications to manage the streams to the platform.

A web-based management application gives authorized users the ability to create, update, remove, and manage users, projects, and data. This application allows for access to logging data, and to source data so that administrative functions are available to general administrators and system admins.

A REST API was created that allows for new tools to be created and integrated into the system. Provided proper authentication and privileges, it is able to interact with all elements of the system functionality, enabling developers to further expand the capabilities of the platform environment. This API even can enable federated learning efforts between two deployments of the platform.

\section{Workflow}

The implemented platform, the Medical Record Longitudinal Information AI System, MERLIN, has the following benefits:

\begin{itemize}
    \item \textbf{Project Agnostic Design} Data are stored and can be re-used in multiple projects. The system is adaptable to various needs of a diverse research environment, and need not be rebuilt for each new project.
    \item \textbf{Quality Assurance and Quality Control} This platform is transparent to investigation into data quality and error. It contains methods and tools to evaluate and mitigate issues with data translation and curation.
    \item \textbf{Minimization of Effort for both Implementation and Usage} The system is simple to build, maintain and update, and has easy to use interfaces. This reduces the effort needed to develop and interact with the system.
    \item \textbf{Modular Construction} The modular nature of the platform's design enables its functionality to be easily expanded with the addition of further modules. This design allows research efforts to start quickly; When new ETL modules are created, they can be added to the platform without additional modification or re-processing of data, allowing the cohort of available data sources to be expanded dynamically. Furthermore, new analysis and modelling approaches can be introduced with new modules as well. This allows the platform to grow with the needs of the research group.
    \item \textbf{Required Human Intervention Minimization} Researchers spend minimal time performing tasks such as data translation and tracking, and instead focus efforts on data analysis and problem solving.
\end{itemize}

Due to the modular nature of the system construction, its design is amenable for use with cloud technologies such as Amazon Web Services (AWS), Google Cloud Platform (GCP), and Microsoft Azure to scale with operation needs.

\begin{figure}[!ht]
\centering
\includegraphics[width=0.86\linewidth]{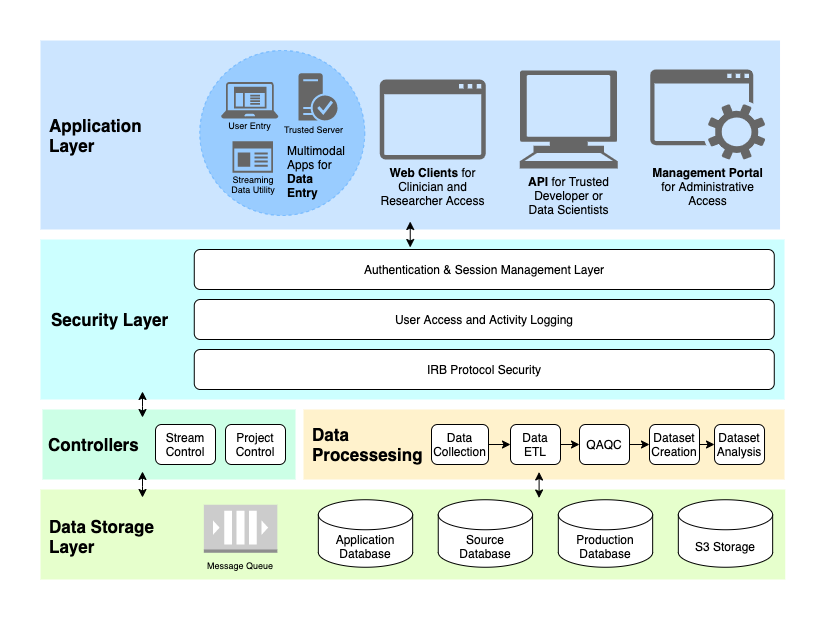}
\caption{An overview of the system architecture for MERLIN.}
\label{fig:merlin_platform}
\end{figure}

Figure \ref{fig:merlin_platform} demonstrates a high-level overview of the implemented system architecture. This system follows the traditional design of a web application, with a user-facing application layer, a security layer, data controllers and data processors serving as the logic layer, and the data storage layer.

\begin{figure}[!ht]
\centering
\includegraphics[width=0.86\textwidth]{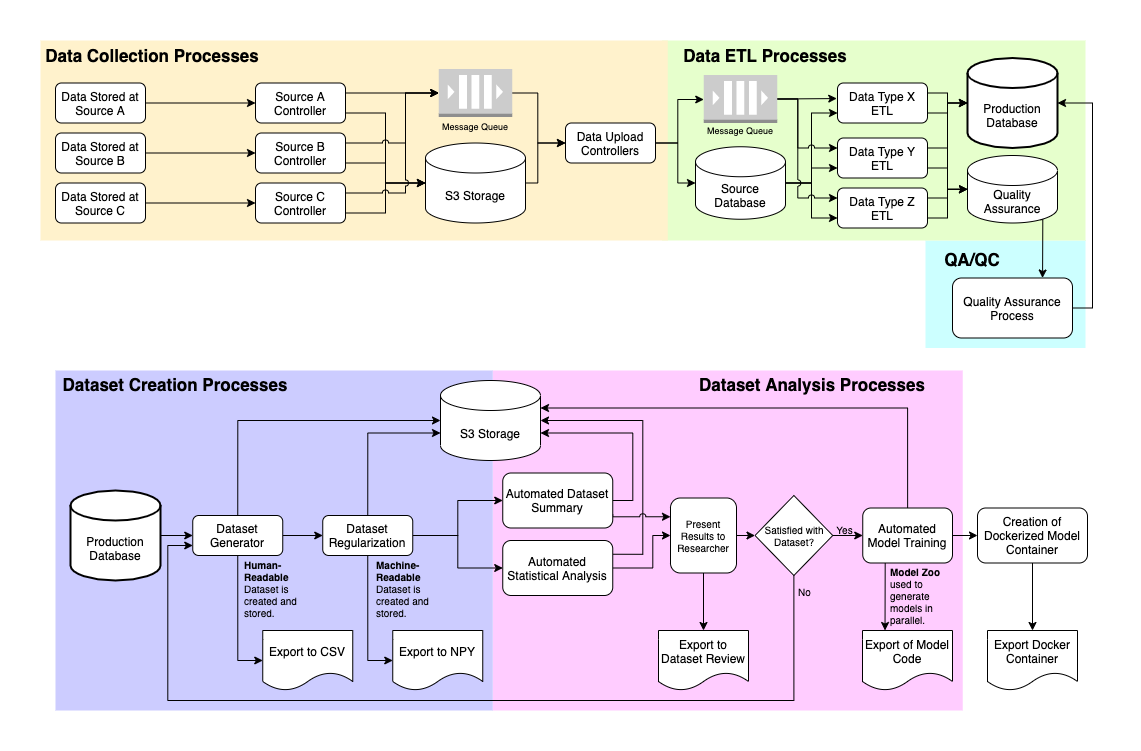}
\caption{The processes involved with the data lifecycle within MERLIN. This includes data collection, ETL, QA/QC, dataset creation and analysis.}
\label{fig:data_pipeline}
\end{figure}

An expanded view of the Data Processing core layer is shown in Figure \ref{fig:data_pipeline}, separated into the five stages addressed in Section \ref{methods}: data collection, ETL, quality assurance and quality control, dataset creation, and dataset analysis. Modules for these stages are highly independent of the user experience in attached applications; instead that is left to the interfaces created to interact with and control the system.

Of those interfaces, the web client portal serves as the main application to access and interact with projects and data. Using it, researchers can pull data from the EHR; interact with the QA/QC pipeline; create and manage projects; design, pull, and analyze datasets, track model progress and training, manage people, data, and projects under IRB protocols; and conduct system auditing all in one place. This portal is designed to simplify the user experience, allowing them to spend their time interacting with data instead of managing computers and data cleaning.

The platform's performance is not specialized in one area. Instead, it serves as a workbench that enables clinicians and data scientists to create cohorts, define phenotypes, and analyze data in methods amenable to the needs of specific projects. This is because the platform stores data in atomic forms, so that they can be combined on-the-fly to suit the needs of research efforts. This allows the inclusion of new variables into models in mere minutes. This even has the benefit that suboptimal results can be detected early, and mitigations, such as thoughtful changes to cohorts' specifications or variables, can be implemented.

MERLIN has been used to-date on multiple projects, which took advantage of components as they became available throughout its development. The system has parsed the medical records of over 200,000 patients from 1989 to present day.

\section{Evaluation}
\subsection{Method}
The platform was benchmarked using a simulated patient data resource. For each data type that the platform ingests, a generator function was written to create random strings that looked like real data with the statistical properties of real patient cohorts. This synthetic resource could be queried like a real EHR resource. The records of 79,143 patients were used to calculate the mean and standard deviation of the size of each string, so that the generator would produce data with this mean and standard deviation. For each category, entries were sorted from most common to least common, and the generator drew from these generated lists of words to create new random entries. This ensured that the ETL modules are benchmarked at their data processing rate, instead of throwing bad and malformed data entries to the QA/QC process, which would have given a false sense of platform efficiency.

To evaluate Objective 1 (Data Acquisition and Ingestion), the platform was benchmarked by running synthetically generated data from a test-EHR system through the intake pipeline. This was done for varying numbers of patients: 1, 5, 10, 50, 100, 500, and 1000 to evaluate how the system can scale with simultaneous load of patients to be ingested. Each patient was simulated with between 2000 and 5000 entries representing approximately 8.73GB of data for the 1000 patient test. For this test, the number of working modules was held constant, and the time until completion, CPU usage, and Memory usage was reported. The objective is that the system can perform in close to linear computational time. A second test was performed in which the number of patients was held constant at 1000, and instead the number of nodes was varied between 1 and 3 to observe how the platform's performance changes with horizontal scaling. Each node contains 100 worker modules. The time until completion, CPU usage per node, and Memory usage per node was reported. The objective is that the system will decrease each of these metrics with additional nodes. The measured operation is a one-time ingestion cost per patient.

To evaluate Objective 2, the platform was benchmarked for the creation of datasets. Although benchmarks attempt to capture the system performance, this step is highly use-case dependent, and is influenced by the complexity of the query. The test scaled the number of variables and measured the response time. The variables added were held constant at variables requiring some computation (not the identity function) and not high degrees of custom computation to give a sense of the expected average computation time. This test was performed for 1, 5, 10, 50, 100, 500 and 1000 unique variables. Response time, CPU and Memory usage were measured. This is not a one-time cost, and occurs each time a user requests a new dataset. Therefore, it is far more important that this metric is lower than the tests evaluating Objective 1.

To evaluate Objective 3, the dataset summary and statistics packages within the platform were benchmarked. While this result is highly variable between use-cases, a success in this test demonstrates that the data is readily available for use in analysis packages. This test was performed for 1, 5, 10, 50, 100, 500 and 1000 unique variables, using the datasets generated in the previous test. Response time, CPU and Memory usage were measured.

\subsection{Results}
\begin{figure}[!h]
\centering
\begin{subfigure}[t]{0.29\linewidth}
\centering
\includegraphics[width=\linewidth]{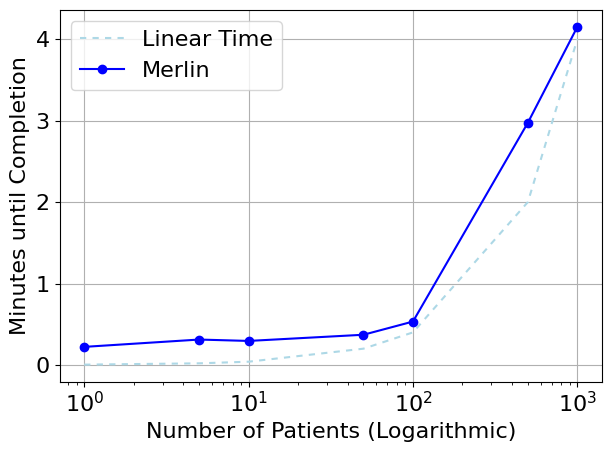} 
\caption{Runtime}
\label{fig:single_node_time}
\end{subfigure}
\begin{subfigure}[t]{0.29\linewidth}
\centering
\includegraphics[width=\linewidth]{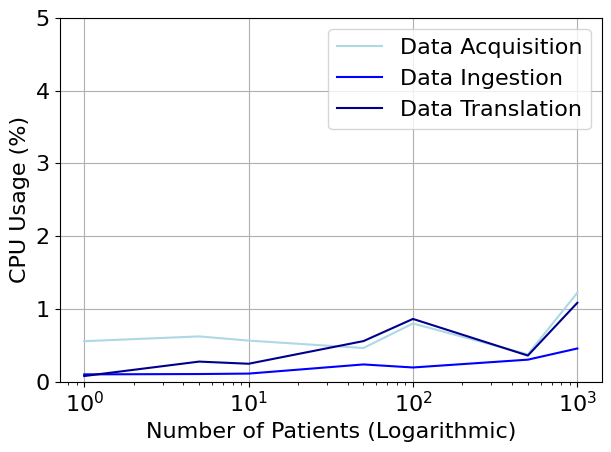} 
\caption{CPU Usage}
\label{fig:single_node_cpu}
\end{subfigure}
\begin{subfigure}[t]{0.29\linewidth}
\centering
\includegraphics[width=\linewidth]{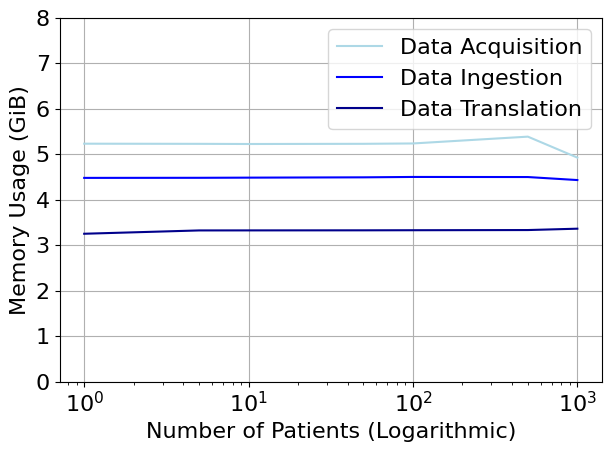}
\caption{Memory Usage}
\label{fig:single_node_mem}
\end{subfigure}

\caption{Benchmark of the Ingestion Process for across a single node. In this test, the number of synthetic patients is varied while the number of compute nodes is held constant.}
\label{fig:singlenode_test}
\end{figure}

Testing used up to three machines (nodes) with 8-Core 2.3Ghz CPUs, and 6GB of allocated RAM all throttled with a network speed of 100Mbps both for upload and download. Each node had 100 worker modules running. Figure \ref{fig:singlenode_test} shows the dataset ingestion test scaling up the number of patients to ingest for a single node. Figure \ref{fig:single_node_time} shows that the process finishes in linear time, with 1000 patients processed in just over 4 minutes. This ingestion and translation process is a one-time cost in the system, and after, those patient records can be used repeatedly on multiple projects. Figure \ref{fig:single_node_cpu} and Figure \ref{fig:single_node_mem} show that the system is linear in CPU and constant in memory respectively.

\begin{figure}[!h]
\centering
\begin{subfigure}[t]{0.29\linewidth}
\centering
\includegraphics[width=\linewidth]{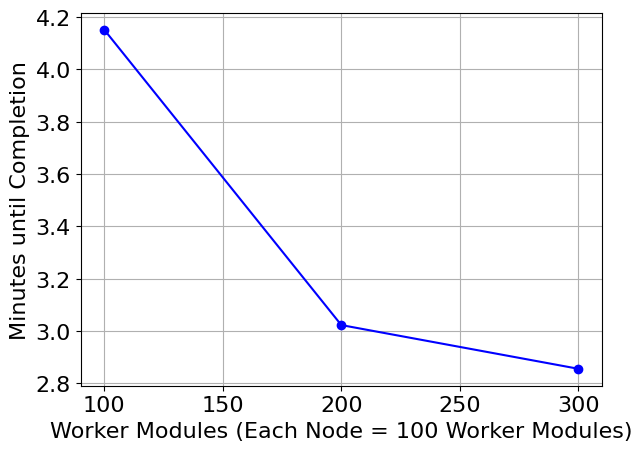} 
\caption{Runtime}
\label{fig:many_nodes_time}
\end{subfigure}
\begin{subfigure}[t]{0.29\linewidth}
\centering
\includegraphics[width=\linewidth]{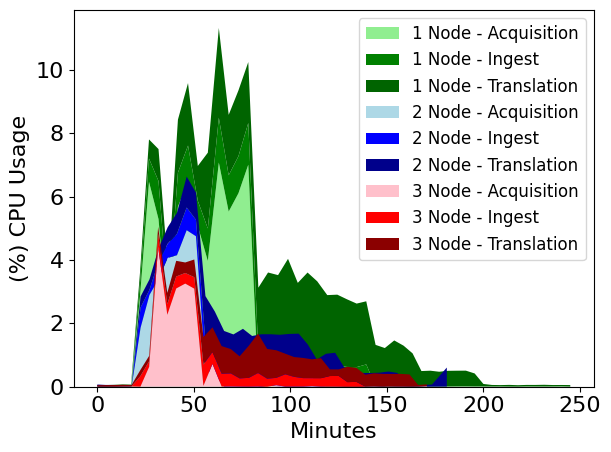} 
\caption{CPU Usage}
\label{fig:many_nodes_cpu}
\end{subfigure}
\begin{subfigure}[t]{0.29\linewidth}
\centering
\includegraphics[width=\linewidth]{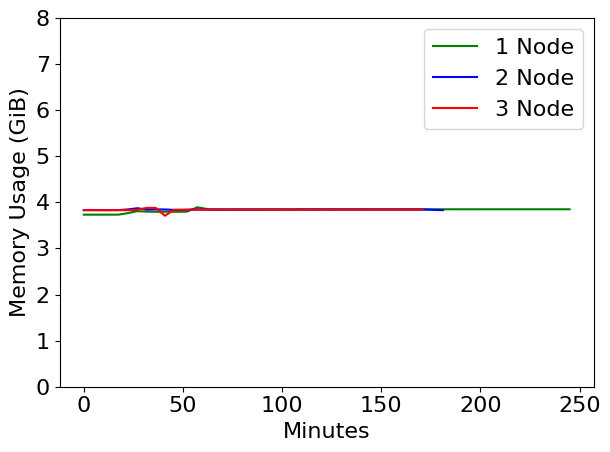}
\caption{Memory Usage}
\label{fig:many_nodes_mem}
\end{subfigure}

\caption{Benchmark of the Ingestion Process for across multiple nodes. In this test, the number of synthetic patients is held constant at 1000 patients.}
\label{fig:multinode_test}
\end{figure}

Figure \ref{fig:multinode_test} illustrates the horizontal scalability of the system, with the number of ingested patients held constant at 1000. Figure \ref{fig:many_nodes_time} shows a decrease with increased nodes. However the long tails in Figure \ref{fig:many_nodes_cpu} indicate that the process of ingesting and translating data are drawn out further than acquisition. This can be explained by a rate limit on write operations for the database resource that the platform was provisioned, but the expectation is that when unconstrained, these processes will benefit from a decreased run-time. Like the previous test, Figure \ref{fig:many_nodes_mem} shows that the system is constant for memory, consistent with the design as stateless services.

MERLIN, in its benchmarked deployment, is able to process approximately 15,000 patients per hour. By comparison, typical processes can take weeks to months. MERLIN's processing capacity can be expanded with increased resources.

These tests suggest that previously stated Objective 1 has been accomplished for the needs of a general use-case. In these tests, a longer ingestion time for a large number of patients is acceptable because it is a one-time cost, whereas the subsequent steps should happen faster.

\clearpage

\begin{figure}[!h]
\centering
\begin{subfigure}[t]{0.29\linewidth}
\centering
\includegraphics[width=\linewidth]{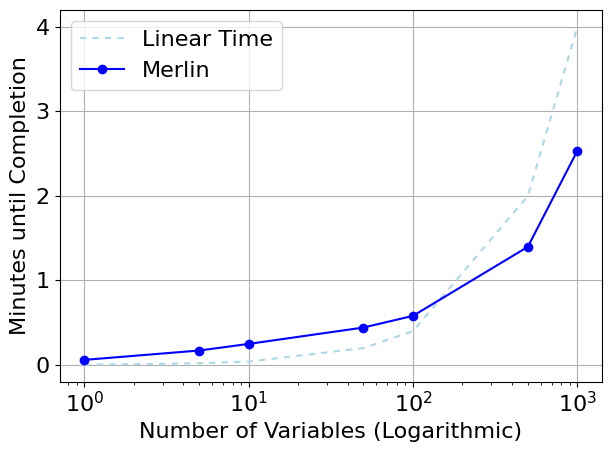} 
\caption{Runtime}
\label{fig:dataset_generation_time}
\end{subfigure}
\begin{subfigure}[t]{0.29\linewidth}
\centering
\includegraphics[width=\linewidth]{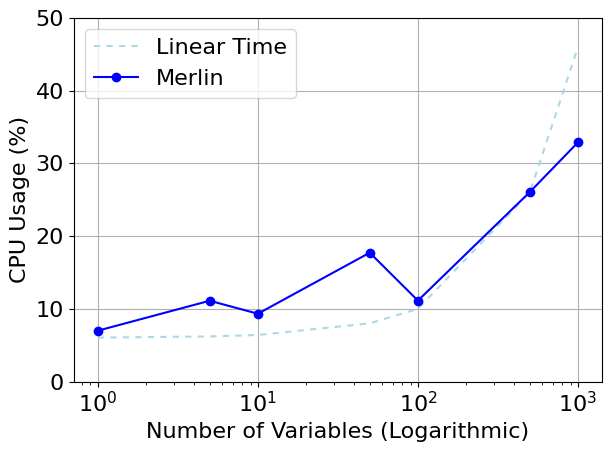} 
\caption{CPU Usage}
\label{fig:dataset_generation_cpu}
\end{subfigure}
\begin{subfigure}[t]{0.29\linewidth}
\centering
\includegraphics[width=\linewidth]{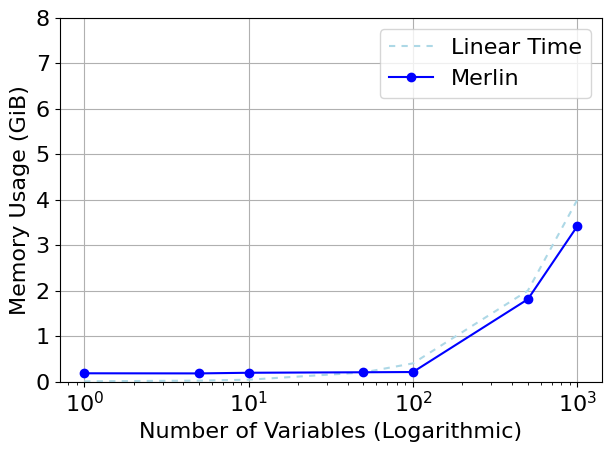}
\caption{Memory Usage}
\label{fig:dataset_generation_mem}
\end{subfigure}

\caption{Benchmark of the Dataset Generation process.}
\label{fig:dataset_generation_test}
\end{figure}

\begin{figure}[!h]
\centering
\begin{subfigure}[t]{0.29\linewidth}
\centering
\includegraphics[width=\linewidth]{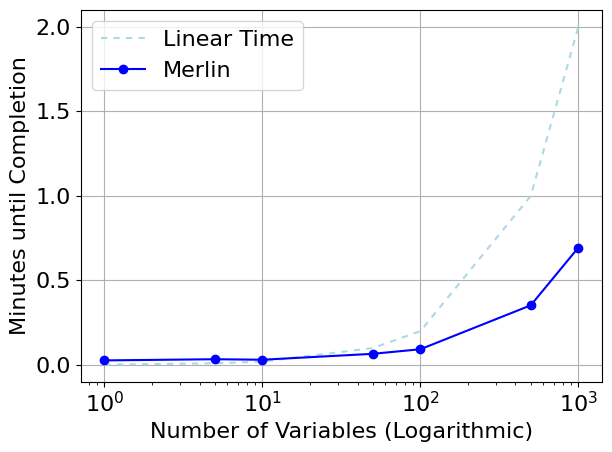} 
\caption{Runtime}
\label{fig:dataset_analysis_time}
\end{subfigure}
\begin{subfigure}[t]{0.29\linewidth}
\centering
\includegraphics[width=\linewidth]{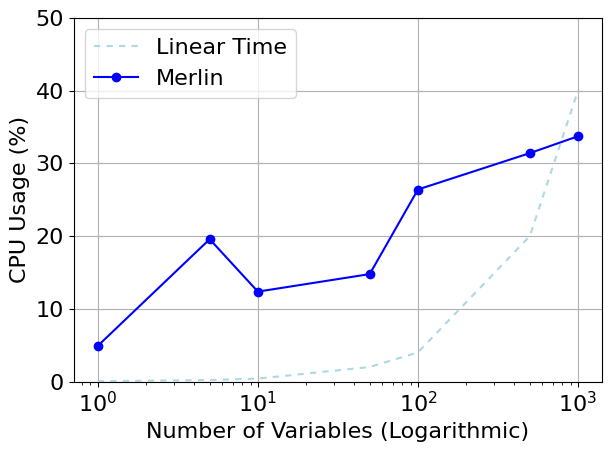} 
\caption{CPU Usage}
\label{fig:dataset_analysis_cpu}
\end{subfigure}
\begin{subfigure}[t]{0.29\linewidth}
\centering
\includegraphics[width=\linewidth]{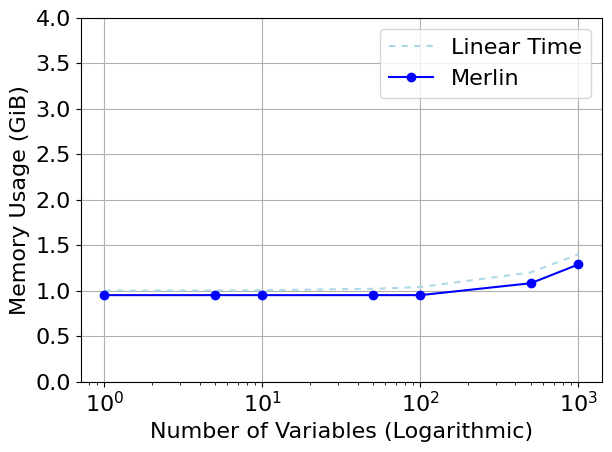}
\caption{Memory Usage}
\label{fig:dataset_analysis_mem}
\end{subfigure}

\caption{Benchmark of the Dataset Analysis process.}
\label{fig:dataset_analysis_test}
\end{figure}

\clearpage

This is indeed what is seen in Figure \ref{fig:dataset_generation_test} and Figure \ref{fig:dataset_analysis_test}. Figure \ref{fig:dataset_generation_test} benchmarks the dataset generation process. In this test, the number of computed variables is scaled up. The test was conducted holding the number of output rows constant, 100,000 synthetic entries across 80,000 synthetic patients in each test (simulating 1.25 events of interest per patient), despite increasing number of variables. The runtime, CPU and Memory usage all scale linearly. With 100 computed variables, the generation time is only slightly over 30 seconds. Indeed even with 1000 computed variables, this process takes 2.5 minutes. By comparison, this task could otherwise take days to weeks to properly regularize data and build custom pipelines for each project, dependent on data formats and quality. This suggests that Objective 2 has been accomplished.

Figure \ref{fig:dataset_generation_test} shows the benchmark test for the preliminary summary and statistics packages to run in the analysis pipeline. This confirms that the datasets generated are indeed readily available for analysis, which is sensible given that they are stored as CSVs with numerical representation or NPY files. This test shows that runtime, CPU, and Memory all scale linearly, although the CPU usage is subject to fluctuation based on the content of the dataset. Even for 1000 computed variables, this process took only 45 seconds, and for 100 computed variables, a mere 7 seconds. This suggests that Objective 3 has been accomplished.

\section{Discussion}
MERLIN enables interactive and iterative research pipelines utilizing data from different sources such as EHRs, medical devices, and prospectively collected data. The short turnaround times for dataset generation and analysis mean that cohorts can be developed and analyzed in near real-time. This platform has the further benefit that it is not specialized to a single task or workflow, which might necessitate rebuilding it for each new project. Instead, adapters can be developed to add new data. The standardized nature of data storage means that even with the inclusion of new data sources, the dataset generation and analysis workflows do not require any change to accommodate the increased data available.  Likewise, if the dataset generation functions or analysis modalities are improved, data do not need to be re-imported or processed to take advantage of new capabilities. Our initiative aimed to create a platform that is generalized to the needs of a biomedical data research team. This framework is built upon standardized components that can be copied as templates and used to expand the included functionality rapidly. 

Other solutions have benefited from NLP-based models to generate complex phenotypes \cite{2018UsingMLtoIdentify}. Although the framework allows for its inclusion, this is notably absent from our current implementation. These NLP models could serve as the inner function of ETL modules referenced in Section \ref{etlmodules}, so long as the output predictions from these models obey the outlined principles for data representation. Furthermore, the included logging and data linking would allow quick references between NLP models and data predicted by those models. In this fashion, NLP models could be incorporated in our framework to leverage their capabilities. NLP-based models were not included in the current implementation due to their fragility with small perturbations, new concepts, or concept drift, \cite{2020ReviewofChallenges, 2020NLPStressTest}, which creates a more significant burden on the QA process and the institution-specific clinical notes structure. For older long-form and unstructured text data, the inclusion of these models may be a necessity. Still, the number of usable patients gained through this approach may be a minimal addition to the study cohort and not worth the development effort.

In addition to MERLIN’s generalizability over its input, it is both able to scale in large environments and can be performant in resource-limited settings such as academia. This gives it the ability to work within the constraints of small research labs and grow with their needs. Our system is designed to take advantage of available resources, handle more concurrent data streams, and run faster if migrated to larger compute centers. The platform can grow without many bottlenecks encountered in other solutions due to algorithmic modularity and separation of each task. Instead of one long, expensive job as found in monolithic architectures, tasks are divided into hundreds of smaller jobs, which may take advantage of hardware parallelization in a more resource-savvy way. Ultimately, this system may benefit from the infrastructure available in AWS, GCP, and Microsoft Azure. 

MERLIN’s usability is not only due to its efficiency and scalability, but also its attention to interfaces and integration. Key to the rapid and efficient use of the platform is the minimization of barriers in interfaces to collect data, generate datasets, and visualize data during the analysis process.

The system is thus targeted to allow researchers to apply their domain-specific knowledge to problems with near real-time feedback and to do so in a framework that supports many projects in parallel. This can even facilitate tasks such as data ingestion, dataset generation, and data analysis to be completed in hours and days instead of months or even years.

\section{Conclusion}
We created a versatile platform to address diverse needs throughout the healthcare AI research pipeline. It is designed to facilitate data collection from multimodal sources, such as EHRs and live data from devices, into a high-fidelity centralized repository with rigorous quality control in ML-ready formats. This methodology enables the rapid creation, evaluation, and analysis of datasets drawn from this repository with version controlling and logging, which can be directly used to train models, discover relationships in the data and produce results efficiently. We anticipate that this approach will spearhead real-world AI model development, and, in the long run, meaningfully improve healthcare delivery.

\section*{Acknowledgements}
The work presented in this paper has been aided by numerous interns and students over the past two years. Most notably, Jonathan Dong, Braden Eberhard, and Sheridan Rea helped with writing the code. In addition, Dr. Heidi Lightfoot, Logan Scoon, Alexis Guttierez, Hassan Hassan, and Dr. Christopher Mukasa have helped in the multiple stages of data collection, interpretation and translation. This work has been supported in part by funding from Partners Innovation, the Brigham Research Institute, Brigham and Women's Hospital Department of Anesthesiology, and the Connors Center IGNITE Award.

\bibliographystyle{unsrt}
\bibliography{references}

\end{document}